\documentclass[aps,pra,preprint,showpacs,groupedaddress,nofootinbib]{revtex4-1}

\usepackage{graphicx}
\usepackage{dcolumn}
\usepackage{bm}
\usepackage{color}
\usepackage{amsmath}
\usepackage{mdframed}
\usepackage{physymb}
\usepackage{enumitem}

\definecolor{red}{rgb}{1,0,0}
\definecolor{orange}{rgb}{1,0.5,0}
\definecolor{green}{rgb}{0.13,0.55,0.13}{
\definecolor{purple}{rgb}{0.5,0,1}

\newcommand{\am}{Amp\`ere's }
\begin{document}

\title{Unpacking Students' Use of Mathematics in Upper-division Physics{:}\\ {Where do we go from here?}}

\author{Marcos D. \surname{Caballero}}
    \email[Corresponding Author: ]{caballero@pa.msu.edu}
	\affiliation{Department of Physics and Astronomy, Michigan State University, East Lansing, MI 48824}
	\affiliation{CREATE for STEM Institute, Michigan State University, East Lansing, MI 48824}
\author{Bethany R. \surname{Wilcox}}
\affiliation{Department of Physics, University of Colorado Boulder, Boulder CO 80309}
\author{Leanne \surname{Doughty}}
\affiliation{Department of Physics and Astronomy, Michigan State University, East Lansing, MI 48824}
\author{Steven J. \surname{Pollock}}
\affiliation{Department of Physics, University of Colorado Boulder, Boulder CO 80309}

\date{\today}

\begin{abstract}
{In their study of physics beyond the first year of University -- termed {\it upper-division} in the US},  {many} of students' primary learning opportunities come from working long, complex back-of-the-book style problems, and from trying to develop an understanding of the underlying physics through solving such problems. Some of the research at the upper-division focuses on how students use mathematics in these problems, and what challenges students encounter along the way. There are a number of different and diverse {research} studies on students' use of mathematics in the {upper-division. These typically utilize one of two broad approaches}, with some researchers primarily seeking out and addressing challenges students face, and others working chiefly to unpack students' in-the-moment reasoning. In this paper, we  {present and} discuss both approaches, and then  {review} research efforts that strive to connect these two approaches in order to make sense of students' use of mathematics as well as to uncover particular challenges that students encounter. These {recent} efforts represent a small step towards synthesizing the two approaches, which we argue is necessary to more meaningfully impact student learning at the upper-division. We close  {our review and discussion} with suggested refinements {for future research questions} for the {physics education research} community to consider while it works to understand how students use math in upper-division courses.
\end{abstract}

\maketitle

\section{Introduction\label{sec:intro}}
Physics education research {(PER)} has a long history of investigating how students approach solving problems, particularly in {1st year, university physics courses -- termed {\it lower-division} or {\it introductory} in the US} (e.g., \cite{McDermott:1999tz,Hsu:2004kh,Meltzer:2012eg}). More recently, a substantial body of work has developed around the myriad of mathematical and conceptual challenges that 2nd, 3rd, and 4th year {university} students encounter in their {\it upper-division} courses (e.g., \cite{Pepper:2010we,Wallace:2010bb,Singh:2006tv,Deslauriers:2011kn,Smith:2013im,Zhu:2012vo,Sayre:2008gh, PhysRevSTPER.7.020113, PhysRevSTPER.8.010111, PhysRevSTPER.8.023101,Ambrose:2004ed,Smith:2010wx,Pollock:2007wv,Mason:2010jd,Smith2009entropy,Meltzer2009thermal,Michelini2015}).\footnote{The term {\it upper-division} refers to students and courses that are beyond the first year of study in physics. In the US, upper-division courses are often solely populated by students who are seeking a physics (or astronomy) degree where {\it lower-division} courses might be taught to science and engineering students as well as non-science majors. As a result, the enrollment in upper-division courses tends to be at least an order of magnitude smaller than lower-division courses.} Those of us who conduct such work have appropriated much of the theory and methods from introductory physics research. However, student difficulties research in the upper-division presents several challenges that are less present in the lower-division: {\it More sophisticated models} -- Students taking upper-division physics courses are working with canonical (yet conceptually complex) models that form the basis for future study in advanced physics. {\it More complex mathematics} -- In upper-division courses, students must learn to use sophisticated mathematical tools (such as multi-variable integration, and {Fourier expansions}) to grapple with these new physics models. {\it Longer and more complicated problems} -- It is in the upper-division where students learn to solve challenging, multi-step problems, which synthesize the aforementioned physical models and mathematical tools.

As such, upper-division course work often includes working back-of-the-book problems that are more sophisticated, more complex, and longer than those appearing in introductory physics. {Many university instructors argue that} one of the primary learning opportunities {for upper-division students} come when engaging with such problems. Hence, some of the research conducted with upper-division students has focused on how students solve typical back-of-the-book style problems, and often emphasizes how students use mathematics while working these problems (e.g., \cite{10.1063/1.2508697,10.1063/1.2820945,Black:2009tu,Irving:2013bi,PhysRevSTPER.9.020119}). Appropriately, this work emphasizes students' use of mathematics in physics, which has been argued to be different from mathematics alone \cite{Uhden2012,Lopez2015,Michelsen2006,Redfors2013,Karam2014,Redish:2006un}. This body of work attempts to answer three research questions: (1) How do students use and/or reason with mathematics in upper-division physics and how is this shaped over time? (2) What sorts of difficulties do students encounter when using and/or reasoning with mathematics in upper-division contexts and how do these challenges change over time? (3) What sorts of interactions, structures, and artifacts help students to productively use mathematics in upper-division physics? To investigate these questions, researchers have often taken one of two basic approaches: (1) a {\it macroscopic} approach that aims to uncover difficulties that students encounter with particular concepts and tools, and (2) a {\it microscopic} approach that is more grounded in learning theory to understand students' in-the-moment reasoning. Each approach has its own benefits and shortcomings. 

 {This paper has two main purposes: 1) to argue that a more complete understanding of student use of mathematics and greater impact on student learning will come from connecting these two approaches; and 2) to challenge researchers in upper-division physics education with several critical research questions.}
To frame this argument and the resulting research questions, we  {review and synthesize} the existing research on students' use of mathematics in upper-division physics including a sampling of findings thus far (Sec.\ \ref{sec:intersection}).\footnote{We have limited our discussion to findings from research studies at the upper-division level. We have not included discussions about teaching or instruction.}
We then discuss the development and use of an analytical framework \cite{PhysRevSTPER.9.020119}, which has helped us begin to connect our work with work that explicitly leverages learning theory (Sec.\ \ref{sec:framework}). 
By summarizing previous findings (Sec.\ \ref{sec:previous}) and re-interpreting prior work (Sec.\ \ref{sec:interpreting}), we demonstrate the utility of this framework when attempting to organize observed student difficulties into coherent themes. 
 {Interestingly, there are several themes in prior work that our framework captures well and others that it does not.}
We also present the shortcomings of this framework, particularly the need to incorporate it into better-developed theoretical constructs, as a means to argue for a stronger connection between {\it macroscopic} and {\it microscopic} work (Sec.\ \ref{sec:discussion}). 
These arguments lead to a set of more refined open questions for the  {physics education research} community to consider (Sec.\ \ref{sec:nextsteps}).

\section{Investigating the Intersection of Math and Physics\label{sec:intersection}}

Researchers have approached how students use mathematics in upper-division physics from a variety of perspectives. These approaches have been shaped by the evolution of our field including our own appropriation of theory from science education and elsewhere (e.g., resources \cite{10.1063/1.2508697,Black:2009tu}), our development of new theoretical tools (e.g., epistemic framing \cite{Irving:2013bi}), and our particular needs (e.g., research supporting course transformation \cite{PhysRevSTPER.8.010111,PhysRevSTPER.9.020119}). While the literature on how students use mathematics in the upper-division is quite diverse,  {in this review} we have noted two distinct, common approaches that differ in scales, methods, and goals.  {In this section, we provide a common language and definition to denote and to distinguish these two approaches.}

First is an approach where the goal is uncovering specific challenges that students encounter with particular physics concepts or mathematical tools. We will refer to this kind of work as ``{\it macroscopic}'' to distinguish it from work that specifically leverages or develops theoretical frameworks and constructs. This label reflects the scale of analysis for this work, which is typically whole classes. Work of this type has been productive in finding particular challenges that students experience with specific concepts and mathematical tools, and, in some cases, in developing instructional strategies and artifacts that help students negotiate those challenges (see Sec.\ \ref{sec:empirical}). {\it Macroscopic} studies typically use students' written work or interviews to document these challenges. While the research using this approach has produced a substantial base for understanding student difficulties, it has limitations. Findings from different studies are often disconnected from each other and from any overarching structure that would help us understand how student ideas change over time. This approach was appropriated from introductory physics at a time when those working in introductory physics had just begun learning about and using new theoretical tools and methods for their own work.

The second approach leverages theoretical frameworks and constructs about how students learn and reason to generate descriptive accounts of students' use of mathematics using students' own ideas as the basis of these accounts. We will refer to this work as ``{\it microscopic}'' to distinguish it from the former approach. This label is useful because the scale of analysis for this work is typically single students or small groups of students. This work has shown promise in developing rich, descriptive accounts of students' in-the-moment reasoning about mathematics in upper-division physics (see Sec.\ \ref{sec:theoretical}). {\it Microscopic} studies typically use \emph{in situ} observations of students working on problems and semi-structured interviews. While this research has helped develop an understanding of how students use mathematics in physics generally, it has found less use when investigating specific challenges that students encounter with particular physics concepts or mathematical tools. Moreover, how  {microscopic studies} will be used to develop instructional strategies and artifacts has yet to be seen, neither {typically} being {an explicit} goal of this work. {\it Microscopic} studies employ theoretical tools such as resources \cite{Hammer:2000vf}, epistemic games \cite{Collins:1993vz}, and framing \cite{Hammer:2005th} and use an approach that often focuses on several short episodes to unpack students' in-the-moment reasoning.

Our labels (``{\it macro-}'' and ``{\it micro-}'') are matters of convenience that help us easily refer to the different approaches. We do not mean to suggest that that these two approaches exist in well-defined and discrete camps {nor do we mean them to be value-laden terms}. Moreover, our naming convention is not meant to imply that {\it macroscopic} work does not make use of learning theory nor that {\it microscopic} work cannot impact {or inform} student learning. These labels are {merely} meant to communicate the kinds of studies that have been conducted with the goals, designs, methods, and analyses discussed above. Each has produced a body of literature that makes headway on some of the research questions above. Below, we {review} sample {research} studies and {synthesize} findings from both types.

\subsection{Macroscopic studies and findings\label{sec:empirical}}

Research that we have called {\it macroscopic} owes its origin to work conducted early in the history of physics education research (PER) \cite{Meltzer:2012eg}. Much of the early work done in introductory physics focused on understanding what difficulties students encountered with particular concepts and mathematical tools \cite{arons1997teaching}, and on developing instructional approaches and artifacts to help students negotiate those difficulties. For example, the Physics Education Group at the University of Washington conducted an impressive amount of research to understand students' conceptual reasoning about different physics topics (e.g., \cite{heron2011angmom, Lindsey:2012vj, Ortiz:2005hu}). Much of this research was used to develop the University of Washington Tutorials, which have been shown to help students negotiate conceptual difficulties that they encounter in introductory physics courses \cite{pollock2007longitudinal, finkelstein2005tip}.  

{\it Macroscopic} studies often focus on students' written work, and in some cases, coordinate students' written work with interview studies to infer student reasoning.  For example, Smith {\it et al.} \cite{Smith:2013im} used students' written responses to pre-surveys along with both teaching and clinical interviews with a subset of the students to investigate students' understanding of Taylor series in the context of thermal physics.  In this study, students' written work was analyzed to identify descriptive (rather than interpretive) themes.  Interviews were analyzed as case studies with the goal of both description and interpretation of students' work.  A number of studies have used similar mixed methods \cite{PhysRevSTPER.9.020119, PhysRevSTPER.8.010111, Pepper:2010we, Wallace:2010bb, Smith:2010wx, sadaghiani2005qm}.  Others have focused only on students' written work \cite{loverude2009prob, loverude2013mm, bucy2006pDerivs, thompson2005pDerivs, Pollock:2007wv}.  

In reviewing the {\it macroscopic} literature, several overlapping themes in student use of mathematics in the upper-division stand out to us.  Our first theme suggests that, despite the claim from some physics instructors that ``the students just don't know the math,'' fluency with procedural mathematics is often not the primary barrier to student success.  {This fits with the models and frameworks developed by physics education researchers who have argued that using mathematics in physics is quite different to simply using mathematics \cite{Uhden2012,Lopez2015,Michelsen2006,Redfors2013,Karam2014}.} There are a number of instances in the literature where students demonstrate that they are quite successful manipulating and applying mathematical formulas but struggle to interpret or to generate these formulas \cite{Smith:2013im, Smith:2010wx, loverude2009prob, PhysRevSTPER.8.010111}.  Which leads us to our second theme: students often struggle to interpret/make sense of mathematical expressions in terms of the appropriate physics (i.e., connecting the math and physics) \cite{loverude2013mm, PhysRevSTPER.8.010111, Wallace:2010bb, Redish:2006un, thompson2005pDerivs,Michelini2015}.  For example, Loverude \cite{loverude2009prob} looked at the extent to which students could apply the binomial formula to determine probabilities in simple systems in the context of statistical mechanics.  He found that, after tutorial instruction, students were able to apply the binomial formula to determine multiplicities but struggled to correctly interpret the results or determine when the formula was appropriate.  

A third, related theme from the {\it macroscopic} literature is that, while students see many of the mathematical tools and techniques used in upper-division physics in their math courses, the operationalization of these tools in their physics courses can be strikingly different \cite{Redish:2006un, Pollock:2007wv, bucy2006pDerivs}.  For example, physicists often do not explicitly express the functional dependence of variables in mathematical formulas (e.g., representing volume as just `$V$' rather than `$V(P,T)$').  This convention can be a barrier to students who are still learning to connect the context-dependent physical interpretation of a variable with its natural functional dependence.  {E.~Pollock} {\it et al.} \cite{Pollock:2007wv} and Bucy {\it et al.} \cite{bucy2006pDerivs} have documented manifestations of this difficulty in the context of line integrals and partial derivatives in thermodynamics.  

We do not claim that these three themes are exhaustive nor that they represent a consensus from researchers who work in this realm.  We also cannot divorce our identification of these themes from our work on developing the analytical framework that will be described later (Sec.\ \ref{sec:ACER}).  {Common threads like those we have identified are present in the {\it macroscopic} literature but have rarely been  {discussed explicitly}, in part because identifying these themes requires synthesis across research studies and topical areas}. It is, in part, because of this lack of synthesis that we argue there is a need for an overarching organizational structure that can be used to design and to interpret these studies but is also consistent with and grounded in theory on how students use mathematics in physics.  

\subsection{Microscopic studies and findings\label{sec:theoretical}}

Research that we have called {\it microscopic} has its origins in the science and math education work conducted in {primary and secondary schools (termed the {\it K-12} in the US)} (e.g. \cite{Louca2004,Elby2001b}). Recently, some physics education researchers have started to leverage theoretical frameworks and constructs to understand how students develop, access, and use knowledge in introductory (e.g. \cite{Tuminaro:2007hr,Scherr:2009gm,Elby2001a}) and upper-division (e.g. \cite{Sabella:2007ff,Bing:2009gq,Kustusch:2014gz,10.1063/1.2820945,Sayre:2008gh}) physics problem solving. The focus has generally been on students' in-the-moment reasoning when working in groups.

Much of this research is grounded in resource theory \cite{Redish2004,Hammer:2005th}, which outlines the idea that knowledge exists in discrete pieces or sets of consistently associated discrete pieces. Students activate and connect these knowledge pieces (resources) to reach an understanding of concepts and models, and to solve problems. Resources can take many forms, for example: basic ideas about the physical world that students consider obvious (e.g., force causes motion) identified by diSessa as phenomenological primitives \cite{diSessa1988}, symbolic forms where particular conceptual ideas are associated with a certain arrangement of symbols in an equation \cite{Sherin:2001wq}, and procedural resources like \textit{find value} and \textit{choose limits} that may be activated when integrating an expression \cite{10.1063/1.2820945}.

{\it Microscopic} work has found that some resources appear solid while others are more plastic. Sayre and Wittmann \cite{Sayre:2008gh} have examined the generation and development of resources in intermediate mechanics { -- a course taken by 2nd year university physics students}. They utilized a plasticity continuum spanning more solid to more plastic resources, where solid resources are durable, have connections to many other resources, and are unlikely to change. By contrast, plastic resources are less stable in structure and are less likely to be activated in new situations. In their study, some students continued to use Cartesian coordinates where polar coordinates were more appropriate despite having previously demonstrated knowledge of polar coordinates, indicating that Cartesian coordinates are a solid resource for those students. In addition, students needed to derive the details of the polar coordinate system whenever they used it, providing evidence that this is a plastic resource.

Other {\it microscopic} work has described how certain resources are linked. Through interviews with physics majors and graduate students where the students completed a problem requiring them to link force and energy knowledge, Sabella and Redish \cite{Sabella:2007ff} found that for different students resources are connected in different ways. Also, even students who demonstrated having strong associations between force and energy resources separately showed little evidence of linking these sets of conceptual resources in the given context.  To better understand these findings, Tuminaro and Redish \cite{Tuminaro:2007hr} adapted the epistemic games construct \cite{Collins:1993vz} to describe the organizational structure of associated procedural resources that they observed students using during introductory physics problem solving. They identified six epistemic games, each with their own specific entry and exit conditions and moves that occur in a unique linear order. {These games are broad and are consistent across different physics content.} Research at the upper-division has shown that although the games played by students in more advanced problem solving have similar moves to {the games that introductory students play}, {a narrower scope is often required to describe the type and order of activated resources} in longer and more complex procedure-oriented problems \cite{10.1063/1.2820945,Kustusch:2014gz}. For example, {\it Finding a Family of Functions} and {\it Fitting the Physical Situation} games are facets of the larger {\it Mapping Meaning to Mathematics} game that students might play when solving a first order differential equation \cite{10.1063/1.2820945}.

{\it Microscopic} work has found that the way in which students frame an activity determines what resources are activated and what games they decide to play. If students expect a problem to involve mathematical manipulations to solve, framing it as a quantitative problem, they will use a different set of resources to answer than if they interpret the problem as a sense-making activity \cite{Hammer:2005th}.  As framing is generally tacit, researchers use different means to infer students' expectations during an activity including students' negotiations about the appropriate approach to a problem \cite{Hammer:2005th}, differing justifications in students' reasoning \cite{Bing:2009gq}, characteristics of student discourse \cite{Irving:2013bi}, and student behaviors \cite{Scherr:2009gm}. Research in upper-division courses shows that frame shifts are frequent in student discussions, and that instructor prompts that require limited-time interactions can transition students into more productive frames \cite{Irving:2013bi}.

Our  {review} of  {\it microscopic} studies does not represent all of the theoretical frameworks and constructs that can and have been used in PER, but serves to demonstrate how this particular set of constructs has afforded the PER community a deeper insight into how students construct and use knowledge. Thus far, the  {\it microscopic} studies have focused on refining and developing our understanding of theoretical frameworks within physics contexts and on identifying a variety of useful  {ways} for understanding students' in-the-moment reasoning. Future research that blends these findings with  {\it macroscopic} research can help provide a more complete interpretation of student difficulties and approaches to problem solving, and inform more effective instructional strategies.

\section{Organizing student difficulties\label{sec:framework}}

The Physics Education Research group at the University of Colorado Boulder (PER@C) has conducted education research in courses for {upper-level} physics majors since 2006. This research has been used to inform the transformation of upper-division courses to more student-centric learning environments \cite{Chasteen:2009wn}. Much of the work conducted by current and former members of PER@C in the upper-division has been {\it macroscopic}. Our research has informed course transformation efforts, which includes developing new course materials \cite{2012arXiv1207.6040B, goldhaber2009qmat, Chasteen:2009wn,Pollock:2012uy} and enhancing instructor teaching practice \cite{Chasteen:2012vb}.  As part of these transformation efforts, we have compiled lists of student difficulties from a variety of sources including informal observations, discussions with traditional disciplinary instructors with experience teaching the courses \cite{2012AIPC.1413..291P}, and more formal research efforts \cite{Wallace:2010bb, Pepper:2010we, PhysRevSTPER.9.020119,  PhysRevSTPER.8.010111}.  These lists of student difficulties provide actionable implications for instructors and offer starting points from which to develop clicker questions, tutorials, and other course activities.

While this research has helped us to transform {some} of the upper-division experience for our physics majors, our lists of student difficulties lack coherence. Although our work has helped articulate common challenges that students in the upper-division face, our findings are often disconnected from one another or any overarching structure that would help us understand how student ideas connect and change over time. {This understanding} would strengthen efforts to transform the physics curriculum as a whole rather than single courses (or even single topics) both in terms of informing development of coherent curricular materials and by providing evidence of how we impact student understanding positively through coherent efforts across our courses.

In an attempt to address this lack of coherence, we sought to organize these ideas; however, canonical problems in the upper-division are long and complex, involving the coordination of a number of resources and tools. Student reasoning about the physics and mathematics in these problems is similarly long and complex. To help deal with complicated, upper-division problem solving, we developed an analytical framework called ``ACER'' that has helped us organize our observations and make a small step towards connecting {\it macroscopic} and {\it microscopic} work. We are not arguing that we have completely synthesized {\it macroscopic} and {\it microscopic} efforts, but that ACER is helping us work from one approach (the approach that we used historically) towards the other.  We have found ACER to be a useful tool for organizing observed student difficulties in the context of Taylor series, Coulomb's law, and Dirac delta functions \cite{PhysRevSTPER.9.020119,2014arXiv1407.6311W}, {as well as in on-going work in variety of other areas.} 

\subsection{The ACER Framework\label{sec:ACER}}

The ACER Framework builds on the work of Wright and Williams \cite{wright1986wise}, Heller \cite{1992AmJPh..60..627H}, and Redish \cite{Redish:2006un} to develop an organizing structure for upper-division students' use of mathematics when solving canonical back-of-the-book style problems. The ACER framework was developed through a modified form of task analysis \cite{Catrambone:vf}. Experts solved a series of problems that employed the mathematics in question while reflecting on and documenting their problem-solving process. Discussion and negotiation with other disciplinary experts using a variety of contexts produced the four components of the ACER framework: {\it Activation of the tool} - selecting which mathematical tool will be used to facilitate a solution to the problem; {\it Construction of the model} - developing the appropriate mathematical representation of the problem by mapping the particular physical system onto appropriate mathematical tools; {\it Execution of the mathematics} - completing a series of mathematical operations determined by the choice of tool to develop a solution; {\it Reflection on the results} - determining the quality or reasonableness of the solution by connecting it to prior knowledge or limiting cases. {While it may appear that the ACER framework suggests this process is linear or ordered in some simple way, we argue that the process by which students and experts solve problems is more fluid an non-linear. ACER is not meant to model student work, but rather to act as a framework to organize that work. That is, it is meant to help us seek coherence and commonality among different aspects of using mathematics in physics not to prescribe a particular solution pathway.}

\begin{figure}[t]
\centering
\includegraphics[clip, trim=30mm 50mm 30mm 10mm, width=0.90\linewidth]{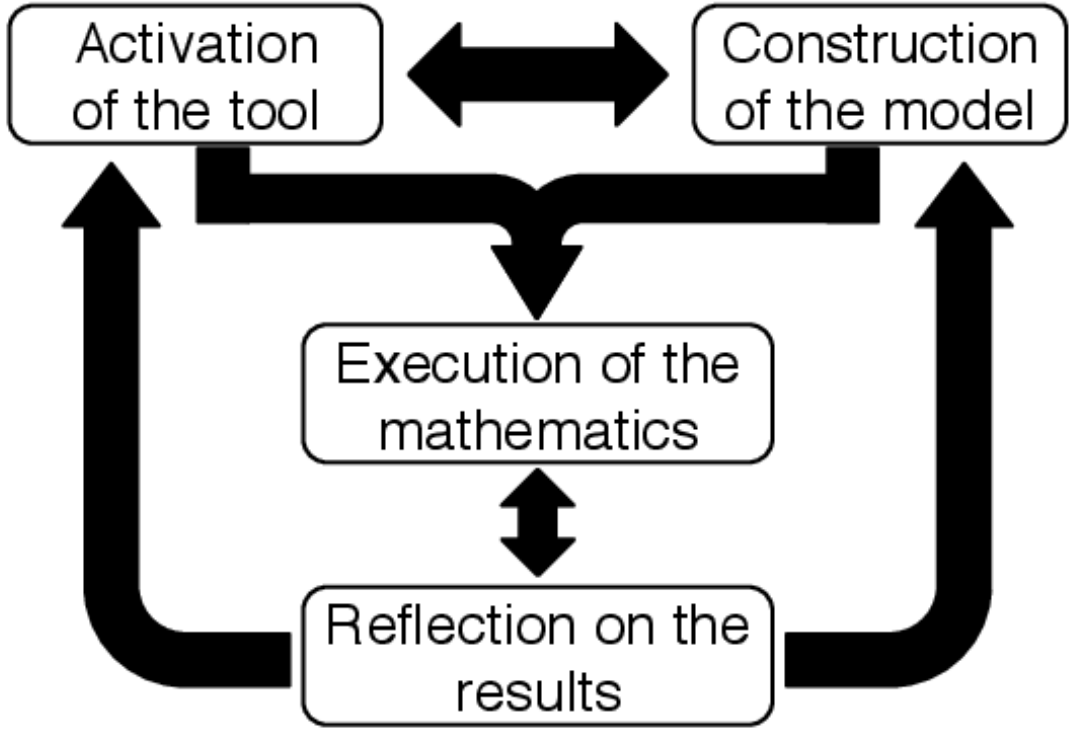}
\caption{A visual representation of the ACER framework. {The arrows connecting the different components of ACER illustrate how experts (during the modified task analysis) moved between different components of the framework.}}\label{fig:acer}
\end{figure}

A visual representation of the framework is shown in Fig.\ \ref{fig:acer}. {The arrows connecting the different components of ACER illustrate how experts (during the modified task analysis) moved between different components of the framework.} Additional details on the design of
the ACER Framework are available in Wilcox, Caballero, Rehn, and Pollock \cite{PhysRevSTPER.9.020119}. The ACER framework has provided an organizational structure for the difficulties we observed students encountering while solving back-of-the-book style problems \cite{Caballero:2012wr, 2012arXiv1207.1283W}. It is an attempt to bring our {\it macroscopic} work closer to theory, but it does not yet explicitly leverage the current theoretical constructs such as epistemic frames \cite{scherr:2009dn} or games \cite{Collins:1993vz}. We will discuss this in more detail later  {(Sec.~\ref{sec:discussion})}.

\subsection{Bridging Approaches}

ACER is an attempt to seek coherence among the myriad of mathematical difficulties that we observed in our course transformation work. For us, it represents a shift away from developing lists of difficulties and toward providing organization and structure to those observations. {ACER is an organizing structure on which we might hang observed student difficulties with mathematics in physics across different contexts.} Through research and development of ACER, we aim to connect the results of {\it macroscopic} work more tightly to learning theory and, thus, to {\it microscopic} work. The design of ACER leveraged learning theory, and the components of ACER are implicitly connected to other theoretical constructs. Although more work is needed to further ground ACER in learning theory, we discuss the design of ACER in the context of the resources framework \cite{Hammer:2000vf} and epistemic frames \cite{Hammer:2005th}.

The resources framework \cite{Hammer:2000vf} underscored the design of ACER. While conducting the task analysis that helped us determine the tacit knowledge that experts used to solve the problems in question \cite{PhysRevSTPER.9.020119}, we unpacked what knowledge (conceptual and procedural) was needed and how we knew to use that knowledge. To make sense of this unpacking, we generated network diagrams that illustrated what ideas were used and how those ideas were organized from our perspective. These diagrams represented an operationalization of the conceptual and procedural resources that we used to solve these problems, and can be thought of as externalized and negotiated resource networks for solving these problems.

The components of the ACER framework (Activation, Construction, Execution, and Reflection) were also developed from the task analysis. While trying to organize the network diagrams into sensible chunks, researchers asked questions like, ``What am I doing here? What is the goal of this part?'' Such questions are epistemic in nature and reflect the history of such frameworks \cite{wright1986wise,1992AmJPh..60..627H,Redish:2006un}. The components of ACER appear to {capture} (at least from the perspective of experts) different epistemic frames. In each component, the expectations are different and these expectations help drive the actions one takes. This idea needs to be investigated more deeply with new studies that focus on clearly defining such frames using {\it in situ} observations of students as well as studying how students shift into different frames \cite{Irving:2013bi}.

ACER was our attempt to connect {\it microscopic} work {and learning theory to our own} {\it macroscopic} work: however, it represents a small step because ACER is a compromise. While the goal was to connect {\it macroscopic} and {\it microscopic} approaches in the hopes of enhancing the work by both, we were sensitive to the need for the ACER framework to be useful for physics instructors who are not grounded in learning theory. We want instructors to {be able to} use this framework to both analyze student work and to design activities that provide opportunities for students to engage in important aspects of solving problems. As we will discuss (Sec.~\ref{sec:discussion}), additional work is needed to build ACER out and connect it more strongly to {\it microscopic} work and learning theory, but we also aim to maintain its instructor-facing utility.

\section{Using the ACER Framework}

{We conducted several studies where the ACER framework was operationalized for specific mathematical tools (Taylor series, Coulomb's Law, Dirac delta functions) \cite{Caballero:2012wr,2012arXiv1207.1283W,2014arXiv1407.6311W}. In this section, we  {review} those studies (post-ACER work). In previous work by our group{, we} did not have the ACER framework to shape our investigations. We also present {modest} reinterpretations of prior work (pre-ACER work) that points out both strengths and shortcomings of using the ACER framework in its current form.}

\subsection{ {A review of findings} from post-ACER work\label{sec:previous}}

The ACER framework was developed while we conducted investigations of student difficulties with Taylor series in {2nd year} classical mechanics \cite{Caballero:2012wr} and Coulomb's Law in {3rd year} electrostatics \cite{2012arXiv1207.1283W}.\footnote{These courses were taught using John R. Taylor's {\it Classical Mechanics} and David J. Griffith's {\it Introduction to Electrodynamics} respectively.}  In addition to the these two studies, we have also used the ACER framework {to} investigate student difficulties with the Dirac delta function in {3rd year} electrostatics \cite{2014arXiv1407.6311W}.   For each of these studies, a task analysis was executed to uncover the conceptual and procedural resources that underlie each ACER component.  We examined students' solutions to exam problems and coordinated their written work with think-aloud interviews designed to target particular difficulties we observed on exam work.  Students' written work and interviews were coded using elements within each component of the ACER framework as anchors.  This section presents a  {review} and synthesis of key findings of each of these investigations organized by component of the framework.

\emph{Activation of the tool:} Unsurprisingly, student success at activating the appropriate mathematical tool to solve a given problem is strongly dependent on the question prompt.  Of the instructor written exam {questions} in the delta function and Taylor series studies, nearly all directly prompted the students what mathematical tool to use, and the overwhelming majority of the students ($>$80\%) used the appropriate tool.  However, when explicit prompting was removed in think-aloud interviews, both studies found that half or more of the students struggled to activate the necessary mathematical tool (i.e., delta functions or Taylor series).  Students in the Coloumb's Law study were more successful in the Activation component: nearly 75\% correctly identified the Coulomb's Law integral for the electric potential as the appropriate tool on exam questions without explicit prompting.  However, exam questions directly targeting Coulomb's Law tend to come early in the 3rd year electrostatics course when students have not been exposed to many other mathematical tools for calculating the electric potential  {(e.g., solving Laplace's equation)}, and this may account for the greater success in Activation.  

\emph{Construction of the model:} The construction component deals with mapping between the mathematics and the physics {-- termed {\it mathematization} in some literature \cite{Uhden2012,Lopez2015,Karam2014}}.  We have found that our students consistently have more difficulty generating appropriate mathematical expressions from physical descriptions of a problem than recognizing the physical meaning of those same expressions.  For example, only three of eight students in an interview setting were able to use delta functions to express the volume charge density, $\rho(\vec{r})$, of a line charge distribution; however, all of these students were able to recognize that charge density as a line charge distribution when provided the correct mathematical expression.  Similarly, when asked to calculate the electric potential from a uniform disk of charge on a midterm exam, nearly half of our students struggled to produce mathematical expressions for the differential charge element ($dq$) and/or the difference vector ($\vec{\scriptr}=\vec{r}-\vec{r}'$) that were appropriate for the specific question at hand.  While generally more successful at interpreting than generating mathematical expressions, we have observed student difficulties with interpreting mathematical expressions in physics contexts.  For example, in interviews only one of eight {2nd year} physics students acknowledged the need for a natural comparative scale when articulating the range for which an approximate expression generated using a Taylor expansion was `good.' 

\emph{Execution of the mathematics:} In the three studies discussed here, we found that issues with procedural mathematics were rarely the primary barrier to student success in problem solving.  In the Taylor series investigation, almost no students in exams or interviews struggled with the relatively simple derivatives required to formally perform the given Taylor expansions.  Mathematical errors were more common in the delta function and integration investigations, where roughly a quarter of the students made various significant mathematical mistakes (beyond dropping a factor of two or a minus sign).  However, these errors were most often accompanied by additional issues in the Activation or Construction components, and closer analysis of students' work did not reveal specific difficulties exhibited consistently across {our population of} students.  

\emph{Reflection on the Result:} We have found that our upper-division students are far less likely than experts to spontaneously reflect on their solutions.  For example, in the integration study we found that less than 10\% of students on exams and only 2 of 10 interviewees made explicit and spontaneous attempts to reflect on their solutions either by checking units or limiting behavior.  However, when prompted to check their solutions in interviews, we found that {3rd year} students consistently suggested checking both units and limiting behavior.  In the Taylor series study, we found that even when prompted to comment on the physical meaning of different terms in the expansion, roughly three-quarters of our {2nd year} students in both exams and interviews struggled to interpret their solutions meaningfully.  

\subsection{Re-interpreting pre-ACER work\label{sec:interpreting}}

In addition to its use to structure new investigations, we argue that ACER can be useful as a tool for organizing and synthesizing the lists of students' difficulties that have been produced in some {\it macroscopic} studies. There are a number of such studies in the literature (e.g., \cite{Pepper:2010we,Wallace:2010bb,Singh:2006tv,Deslauriers:2011kn,Smith:2013im,Zhu:2012vo,Sayre:2008gh, PhysRevSTPER.7.020113, PhysRevSTPER.8.010111, PhysRevSTPER.8.023101,Ambrose:2004ed,Smith:2010wx,Pollock:2007wv,Mason:2010jd,Smith2009entropy,Meltzer2009thermal}). As an example, this section re-examines the various difficulties identified by \citeauthor{Wallace:2010bb} \cite{Wallace:2010bb} and \citeauthor{PhysRevSTPER.8.010111} \cite{PhysRevSTPER.8.010111} through the lens of the ACER framework. We note that to conduct a complete study of these topics using ACER would require a full operationalization {in the unique context of the topical area under investigation}. {That is not our goal here -- instead, we discuss more generally how ACER can help to} interpret the findings of prior studies and to suggest new areas for investigation, as well as how prior work informs us of shortcomings to the ACER framework.

\subsubsection{Summary of previous findings}

\citeauthor{Wallace:2010bb} investigated how upper-level students solve problems involving \am Law \cite{Wallace:2010bb}. Much of the data collected by the researchers came from students' written responses to questions about \am Law. To investigate their findings further, \citeauthor{Wallace:2010bb} also conducted interviews with students and experts (graduate students and an instructor) where interviewees solved \am Law problems. Their findings fall into two categories: (1) students struggle to connect the enclosed current ($I_{enc}$) to the properties of the magnetic field, and (2) students might not use information about the magnetic field to set up the problem. These observations are further unpacked below \citep[p.~4--7]{Wallace:2010bb}. We use {\bf AL} (for \am Law) and single letter ({\bf a}-{\bf d}) to refer to these observations.

\begin{description}[noitemsep,nolistsep]
\item[ALa] Some students reason that $I_{enc}=0$ implies $\mathbf{B}=0$.
\item[ALb] Some students claim the magnetic field of a solenoid cannot have a radial component because the problem's solution does not depend on the width of the Amp\`erian loop.
\item[ALc] Some students did not choose an Amp\`erian loop based on the direction in which the magnetic field points.
\item[ALd] Some students did not use the fact that the magnetic field is zero outside the solenoid in their calculations. 
\end{description}

\citeauthor{PhysRevSTPER.8.010111} investigated a variety of additional difficulties that students experience in upper-division electricity and magnetism including those related to Gauss's Law, vector calculus, and electric potential \cite{PhysRevSTPER.8.010111}. For this review, we have chosen to discuss the findings related to Gauss's Law  in detail. These researchers used students' written responses to Gauss's Law questions to determine the prevalence of difficulties. They then attempted to triangulate those findings with think-aloud interviews, much in the same way as \citeauthor{Wallace:2010bb} \cite{Wallace:2010bb}. \citeauthor{PhysRevSTPER.8.010111} did not attempt to classify their observations beyond the content area in which the work was conducted, instead they aimed to ``document an initial list of common student difficulties with the goal of providing resources for improvement of instruction and for future research'' \citep[p.~1]{PhysRevSTPER.8.010111}. The researchers make the following observations of student difficulties with Gauss's Law \citep[p.~12]{PhysRevSTPER.8.010111}. We use {\bf GL} (for Gauss's Law) and single letter ({\bf a}-{\bf d}) to refer to these observations.

\begin{description}[noitemsep,nolistsep]
\item[GLa] Students make incorrect inferences about the electric field based on Gauss's law. Some students inferred that the electric field at any point on a Gaussian surface is determined only by the charge enclosed, even in non-symmetric situations.
\item[GLb] Students are unclear in distinguishing flux and electric field.
\item[GLc] Students struggle to articulate complete symmetry arguments. They have particular difficulty with the geometrical symmetry arguments that expert physicists use.
\item[GLd] Students apply Gauss's law when not appropriate. 
\end{description}

\subsubsection{ACER captures some prior observations}

Some of the findings presented by \citeauthor{Wallace:2010bb} \cite{Wallace:2010bb} and \citeauthor{PhysRevSTPER.8.010111} \cite{PhysRevSTPER.8.010111} have clear links to the ACER framework. For example, {a task analysis of \am Law problems would almost} certainly find that, in the Construction component, information about the direction and strength of the magnetic field should be used to determine an appropriate Amp\`erian loop. The observation ({\bf ALc}) that some students ``did not choose their Amp\`erian loop based on the direction in which the magnetic field points'' suggests that these students are not mapping the mathematics of this loop integral onto the physical situation.  This is consistent with findings from our integration study \cite{PhysRevSTPER.9.020119}, which found that some students did not use the geometry of the physical situation when expressing the difference vector, $\scriptr$.  Observation {\bf ALd}, that students ``did not use the fact that the magnetic field is zero in their calculations'' \citep[p.~5--6]{Wallace:2010bb}, is another indication of a difficulty with mapping between the mathematics and the physics (i.e., Construction).

ACER also highlights connections between several of the findings reported by \citeauthor{PhysRevSTPER.8.010111} and our work on student difficulties with Coulomb's Law \cite{PhysRevSTPER.9.020119}. In our discussion of the Activation component for Coulomb's Law \citep[p.~5]{PhysRevSTPER.9.020119}, we presented four elements that are needed to activate its use. One of these is that the symmetry of the problem is not conducive to using Gauss's Law. Thus a task analysis for Gauss's Law would almost certainly find (among other elements) that one must evaluate when the symmetry is appropriate before utilizing Gauss's Law. The findings from \citeauthor{PhysRevSTPER.8.010111} ({\bf GLc} and {\bf GLd}) that ``students struggle to articulate complete symmetry arguments'' and ``students apply Gauss's law when not appropriate'' are both consistent with our observation that some students attempt to use Gauss's Law in problems where Coulomb's Law is appropriate \cite{PhysRevSTPER.9.020119}.  These issues suggest that students struggle with the interplay between Construction and Activation when working with Gauss's Law.

\subsubsection{ACER identifies new area for research}

We identified difficulties presented in \citeauthor{Wallace:2010bb} that align with the Construction component of ACER. However, we are unable to comment on the presence or absence of difficulties associated with Activation, Execution, or Reflection. Their study did not investigate these aspects of the problem-solving process in the context of \am Law. Future studies aimed at capturing other components of ACER might include confronting students with Activation tasks where they must decide whether to use \am Law instead of Biot-Savart, or Reflection tasks that ask them discuss their solutions. The analysis of student work in \citeauthor{Wallace:2010bb} focused on how students set up the problem, which is why we can find elements of a Construction component in their findings.

\subsubsection{ACER does not capture some findings}\label{subsubsec:missing}

In addition to questions asking students to manipulate \am Law, \citeauthor{Wallace:2010bb} asked students to solve more conceptual problems (e.g., {\it In a cylindrical tube where the current is uniformly distributed, where is the magnetic field largest?}).  The observations ({\bf ALa} \& {\bf ALb}) that some students ``reason that $I_{enc}=0$ implies $\mathbf{B}=0$'' and ``claim the magnetic field of a solenoid cannot have a radial component because the problem's solution does not depend on the width of the Amp\`erian loop'' \citep[p.~4]{Wallace:2010bb} are not clearly associated with any components of ACER. {Exclusively} conceptual problems such as those that elicited these difficulties were not the focus of the original ACER design. {ACER was designed to investigate how students use mathematics in physics. While this includes conceptual reasoning, particularly in the Construction and Reflection components of the framework, it does so with the specific intent of using sophisticated mathematical tools to gain insight into the underlying physics. Note that we are not arguing that conceptual and mathematical reasoning should or can be divorced, but rather that currently ACER fails to capture difficulties that appear to be exclusively conceptual in nature.}

Similarly, observations identified by \citeauthor{PhysRevSTPER.8.010111} ({\bf GLa} \& {\bf GLb}) that ``students make incorrect inferences about the relationship of the field at a point and the charge enclosed'' and ``students are unclear in distinguishing flux and the electric field'' are of a different nature than the ACER framework was meant to capture.  ACER was an attempt to organize what students do when solving typical back-of-the-book style problems, and it has never been used to organize students' work with more conceptual upper-division problems. Below, (Sec.~\ref{sec:discussion}), we discuss these shortcomings in more detail and, later, we will present what aspects of upper-division student problem solving our research community should attempt to capture in future work (Sec.~\ref{sec:nextsteps}).

\section{Discussion\label{sec:discussion}}

Our work to transform teaching and learning in the upper-division is {largely} {\it macroscopic}; we leverage theoretical frameworks and constructs in our thinking, development, and study design, but do not yet explicitly ground our work in theory. This is in contrast to some of the excellent work that others have done that is explicitly grounded in learning theory, which we have characterized in this  {review} as {\it microscopic}. The ACER framework was developed to help us seek coherence among the lists of difficulties we observed while transforming teaching and learning at the upper-division. We have argued that it represents (for us) a move to seek a stronger connection to theory. Below, we discuss the merits and shortcomings of this approach.

\subsection{Merits of the ACER Framework}

The merits of the ACER framework stem in part from its generalizable nature. ACER can be used in a variety of contexts because it addresses key elements in problem-solving at the upper-division, and, as such, provides actionable information for instruction. The components of ACER (Activation, Construction, Execution, and Reflection) speak to critical aspects of the process of developing a solution to most back-of-the-book style problems, which represent the bulk of the opportunities for upper-division students to develop their problem-solving practice. Each of these components is present (at least implicitly) in the work needed to solve such problems. The challenges that we observe students experiencing in these different areas provide instructors with information that allows them to design additional course activities for students to engage with those elements. For example, in our work with Taylor series, we observed that students struggle to judge when using Taylor series is appropriate (Activation). We have since developed additional course activities where students discuss when and why Taylor series can be used in particular problems \cite{CUmaterials}.

The flexibility of the ACER framework is another strength because ACER can be operationalized for a wide variety of mathematical tools. Operationalizing the ACER framework relies on conducting a detailed tasks analysis for the mathematical tool, documenting all the steps that one would take, and attempting to make those steps more generalizable while working additional problems. It is here where we leverage the resources framework \cite{Hammer:2005th} to document and to organize the conceptual and procedural resources needed to work exemplar problems. The resulting externalized and negotiated resource network provided opportunities to seek coherence across the use of different mathematical tools. In fact, it was separate task analyses of Taylor's series \cite{Caballero:2012wr} and Coulomb's Law \cite{2012arXiv1207.1283W} that were merged once we realized that overarching structures were quite similar \cite{PhysRevSTPER.9.020119}. We have since applied the ACER framework to Dirac delta functions with similar success \cite{2014arXiv1407.6311W}.

Because of its generalizable nature and flexibility, the ACER framework has helped us find some coherence among the challenges that we observe students exhibiting with particular mathematical tools. In our work, we have found evidence to bolster themes that we observed in the literature (Sec.~\ref{sec:empirical}). The main obstacle for upper-division physics students does not appear to be that they ``just don't know the math'' (Execution), but that they struggle to decide when to use particular mathematical tools (Activation), to map particulars of their problem to the general formalism (Construction), and to decide whether their solution is sufficient (Reflection). In our studies, we have found recurrent challenges with student work such as choosing one mathematical tool over another (e.g., Coulomb's vs Gauss's Law), consistently using the chosen coordinate system (e.g., Cartesian vs. spherical), and thoughtfully judging the validity of a solution (e.g., checking units vs. taking known/useful limits).

Finally, the ACER framework has helped drive investigations into new areas. Our pre-ACER work had not investigated how students choose mathematical tools (Activation) or how students judge their solutions (Reflection) \cite{Wallace:2010bb, PhysRevSTPER.8.010111}. Our previous studies investigated aspects of students' conceptual and mathematical understanding using questions similar to those in post-ACER studies, but they were not specifically designed to interrogate Activation and Reflection. The development of the ACER framework and our early results helped shape studies into these under-researched areas.

\subsection{Shortcomings of the ACER Framework}

The ACER framework represents an attempt to more strongly connect our {\it macroscopic} work to learning theory, and, thus to {\it microscopic} work. It is a conservative attempt that still uses the same data streams that we have collected in the past (i.e., students' written work and think-aloud interviews). Our attempt aims to balance a {better, stronger grounding} in learning theory with a framework that makes sense to {university instructors} who are not as familiar with such work. In our continued research and development of ACER, we aim to adapt the ACER framework to connect more strongly to theoretical constructs such as epistemic frames. However, how such connections should be made is non-trivial and requires working with both approaches to make ACER and these theoretical constructs more consonant. Furthermore, different data streams (e.g., {\it in situ} observations) are needed to investigate students' in-the-moment reasoning. We are in the process of collecting such data and {discussing analysis techniques that can leverage} both approaches.

{Furthermore, that ACER is unable to capture and to help organize some of the conceptual difficulties observed in prior work is unfortunate (Sec.\ \ref{subsubsec:missing}). It is likely that this shortcoming results from the focus of ACER to organize difficulties observed when students are solving typical back-of-the-book style problems. In acknowledging this shortcoming, we are not suggesting that conceptual understanding is not needed for solving such quantitative problems; it certainly is. Instead, we are arguing that the conceptual understanding used when solving quantitative problems, which are representative of upper-division course work, occurs within Construction and Reflection components of ACER. That is, for these kinds of problems, conceptual understanding is often used to construct the model or reflect on the solution. How exclusively conceptual problems fit into ACER framework is not quite clear, because they often require students to develop an explanation or argument, which is not the focus of ACER. This shortcoming will need to be addressed if we are to make better sense of the challenges that students face in upper-division physics.}


\section{Next Steps\label{sec:nextsteps}}

This review provided a  {synthesis} and discussion of the work done on students' use of mathematics in upper-division physics. In doing so, we  {reviewed} the {\it macroscopic} work that we have done at CU-Boulder to understand and to organize the challenges that students face using mathematics on typical back-of-the-book style problems. We have discussed the development and use of the ACER framework, which aims to seek coherence among the challenges we observe in students' written work and in interviews. We have used the ACER framework to re-visit old work with new eyes. Finally, we  {discussed} the merits of our approach as well as its shortcomings. Through this exercise, we reflected on the work that we have done and the work that the community has conducted. We propose a set of research questions that address some of the issues raised in our writing:

\begin{enumerate}[noitemsep,nolistsep]
\item {\it In what ways do students use mathematics across the upper-division physics, and how does that use change over time?}

\item {\it How do difficulties that students exhibit in their work connect to their in-the-moment reasoning about mathematics in physics?}

\item {\it What sorts of instructional tools and strategies can help students come to a deeper understanding of physics through their use of mathematics not simply on individual topics, but across topics, concepts, and courses?}
\end{enumerate}

Some of these questions might be answered through either a macroscopic or a microscopic approach. However, we have argued that a synthetic approach in which researchers blend both approaches and/or collaborate with colleagues across approaches will result in more complete solutions. We believe that the PER community is ready for such a challenge.

\begin{acknowledgments}
The authors would like to thank current and former members of the Physics Education Research groups at CU-Boulder and Michigan State for their work on different aspects of the projects presented in this paper. Particular thanks go to Charles Baily, Stephanie Chasteen, Paul Irving, Rachel Pepper, Kathy Perkins, Daniel Rehn, and Colin Wallace. This work was supported by NSF-CCLI Grant No.~DUE-1023028, the Science Education Initiative, and a National Science Foundation Graduate Research Fellowship under Award No.~DGE 1144083.  
\end{acknowledgments}

\bibliography{math-phys}

\begin{thebibliography}{70}
\expandafter\ifx\csname natexlab\endcsname\relax\def\natexlab#1{#1}\fi
\expandafter\ifx\csname bibnamefont\endcsname\relax
  \def\bibnamefont#1{#1}\fi
\expandafter\ifx\csname bibfnamefont\endcsname\relax
  \def\bibfnamefont#1{#1}\fi
\expandafter\ifx\csname citenamefont\endcsname\relax
  \def\citenamefont#1{#1}\fi
\expandafter\ifx\csname url\endcsname\relax
  \def\url#1{\texttt{#1}}\fi
\expandafter\ifx\csname urlprefix\endcsname\relax\def\urlprefix{URL }\fi
\providecommand{\bibinfo}[2]{#2}
\providecommand{\eprint}[2][]{\url{#2}}

\bibitem[{\citenamefont{McDermott and Redish}(1999)}]{McDermott:1999tz}
\bibinfo{author}{\bibfnamefont{L.~C.} \bibnamefont{McDermott}}
  \bibnamefont{and} \bibinfo{author}{\bibfnamefont{E.}~\bibnamefont{Redish}},
  \emph{\bibinfo{title}{{Resource letter: PER-1: Physics education research}}},
  \bibinfo{journal}{Am. J. Phys.} \textbf{\bibinfo{volume}{67}},
  \bibinfo{pages}{755} (\bibinfo{year}{1999}).

\bibitem[{\citenamefont{Hsu et~al.}(2004)\citenamefont{Hsu, Brewe, Foster, and
  Harper}}]{Hsu:2004kh}
\bibinfo{author}{\bibfnamefont{L.}~\bibnamefont{Hsu}},
  \bibinfo{author}{\bibfnamefont{E.}~\bibnamefont{Brewe}},
  \bibinfo{author}{\bibfnamefont{T.~M.} \bibnamefont{Foster}},
  \bibnamefont{and} \bibinfo{author}{\bibfnamefont{K.~A.}
  \bibnamefont{Harper}}, \emph{\bibinfo{title}{{Resource Letter RPS-1: Research
  in problem solving}}}, \bibinfo{journal}{Am. J. Phys.}
  \textbf{\bibinfo{volume}{72}}, \bibinfo{pages}{1147} (\bibinfo{year}{2004}).

\bibitem[{\citenamefont{Meltzer and Thornton}(2012)}]{Meltzer:2012eg}
\bibinfo{author}{\bibfnamefont{D.~E.} \bibnamefont{Meltzer}} \bibnamefont{and}
  \bibinfo{author}{\bibfnamefont{R.~K.} \bibnamefont{Thornton}},
  \emph{\bibinfo{title}{{Resource Letter ALIP--1: Active-Learning Instruction
  in Physics}}}, \bibinfo{journal}{Am. J. Phys.} \textbf{\bibinfo{volume}{80}},
  \bibinfo{pages}{478} (\bibinfo{year}{2012}).

\bibitem[{\citenamefont{Pepper et~al.}(2010)\citenamefont{Pepper, Chasteen,
  Pollock, and Perkins}}]{Pepper:2010we}
\bibinfo{author}{\bibfnamefont{R.}~\bibnamefont{Pepper}},
  \bibinfo{author}{\bibfnamefont{S.}~\bibnamefont{Chasteen}},
  \bibinfo{author}{\bibfnamefont{S.}~\bibnamefont{Pollock}}, \bibnamefont{and}
  \bibinfo{author}{\bibfnamefont{K.}~\bibnamefont{Perkins}},
  \emph{\bibinfo{title}{{Our best juniors still struggle with Gauss's Law:
  Characterizing their difficulties}}}, \bibinfo{journal}{PERC 2010
  Proceedings} p. \bibinfo{pages}{245} (\bibinfo{year}{2010}).

\bibitem[{\citenamefont{Wallace and Chasteen}(2010)}]{Wallace:2010bb}
\bibinfo{author}{\bibfnamefont{C.}~\bibnamefont{Wallace}} \bibnamefont{and}
  \bibinfo{author}{\bibfnamefont{S.}~\bibnamefont{Chasteen}},
  \emph{\bibinfo{title}{{Upper-division students' difficulties with
  Amp{\`e}re's law}}}, \bibinfo{journal}{Phys. Rev. ST Phys. Educ. Res.}
  \textbf{\bibinfo{volume}{6}}, \bibinfo{pages}{020115} (\bibinfo{year}{2010}).

\bibitem[{\citenamefont{Singh et~al.}(2006)\citenamefont{Singh, Belloni, and
  Christian}}]{Singh:2006tv}
\bibinfo{author}{\bibfnamefont{C.}~\bibnamefont{Singh}},
  \bibinfo{author}{\bibfnamefont{M.}~\bibnamefont{Belloni}}, \bibnamefont{and}
  \bibinfo{author}{\bibfnamefont{W.}~\bibnamefont{Christian}},
  \emph{\bibinfo{title}{{Improving students' understanding of quantum
  mechanics}}}, \bibinfo{journal}{Physics Today} \textbf{\bibinfo{volume}{59}},
  \bibinfo{pages}{43} (\bibinfo{year}{2006}).

\bibitem[{\citenamefont{Deslauriers and Wieman}(2011)}]{Deslauriers:2011kn}
\bibinfo{author}{\bibfnamefont{L.}~\bibnamefont{Deslauriers}} \bibnamefont{and}
  \bibinfo{author}{\bibfnamefont{C.}~\bibnamefont{Wieman}},
  \emph{\bibinfo{title}{{Learning and retention of quantum concepts with
  different teaching methods}}}, \bibinfo{journal}{Phys. Rev. ST Phys. Educ.
  Res.} \textbf{\bibinfo{volume}{7}}, \bibinfo{pages}{010101}
  (\bibinfo{year}{2011}).

\bibitem[{\citenamefont{Smith et~al.}(2013)\citenamefont{Smith, Thompson, and
  Mountcastle}}]{Smith:2013im}
\bibinfo{author}{\bibfnamefont{T.~I.} \bibnamefont{Smith}},
  \bibinfo{author}{\bibfnamefont{J.~R.} \bibnamefont{Thompson}},
  \bibnamefont{and} \bibinfo{author}{\bibfnamefont{D.~B.}
  \bibnamefont{Mountcastle}}, \emph{\bibinfo{title}{{Student understanding of
  Taylor series expansions in statistical mechanics}}}, \bibinfo{journal}{Phys.
  Rev. ST Phys. Educ. Res.} \textbf{\bibinfo{volume}{9}},
  \bibinfo{pages}{020110} (\bibinfo{year}{2013}).

\bibitem[{\citenamefont{Zhu and Singh}(2012)}]{Zhu:2012vo}
\bibinfo{author}{\bibfnamefont{G.}~\bibnamefont{Zhu}} \bibnamefont{and}
  \bibinfo{author}{\bibfnamefont{C.}~\bibnamefont{Singh}},
  \emph{\bibinfo{title}{{Surveying students' understanding of quantum mechanics
  in one spatial dimension}}}, \bibinfo{journal}{Am. J. Phys.}
  \textbf{\bibinfo{volume}{80}}, \bibinfo{pages}{252} (\bibinfo{year}{2012}).

\bibitem[{\citenamefont{Sayre and Wittmann}(2008)}]{Sayre:2008gh}
\bibinfo{author}{\bibfnamefont{E.}~\bibnamefont{Sayre}} \bibnamefont{and}
  \bibinfo{author}{\bibfnamefont{M.}~\bibnamefont{Wittmann}},
  \emph{\bibinfo{title}{{Plasticity of intermediate mechanics students'
  coordinate system choice}}}, \bibinfo{journal}{Phys. Rev. ST Phys. Educ.
  Res.} \textbf{\bibinfo{volume}{4}}, \bibinfo{pages}{020105}
  (\bibinfo{year}{2008}).

\bibitem[{\citenamefont{Ayene et~al.}(2011)\citenamefont{Ayene, Kriek, and
  Damtie}}]{PhysRevSTPER.7.020113}
\bibinfo{author}{\bibfnamefont{M.}~\bibnamefont{Ayene}},
  \bibinfo{author}{\bibfnamefont{J.}~\bibnamefont{Kriek}}, \bibnamefont{and}
  \bibinfo{author}{\bibfnamefont{B.}~\bibnamefont{Damtie}},
  \emph{\bibinfo{title}{Wave-particle duality and uncertainty principle:
  Phenomenographic categories of description of tertiary physics students'
  depictions}}, \bibinfo{journal}{Phys. Rev. ST Phys. Educ. Res.}
  \textbf{\bibinfo{volume}{7}}, \bibinfo{pages}{020113} (\bibinfo{year}{2011}).

\bibitem[{\citenamefont{Pepper et~al.}(2012{\natexlab{a}})\citenamefont{Pepper,
  Chasteen, Pollock, and Perkins}}]{PhysRevSTPER.8.010111}
\bibinfo{author}{\bibfnamefont{R.~E.} \bibnamefont{Pepper}},
  \bibinfo{author}{\bibfnamefont{S.~V.} \bibnamefont{Chasteen}},
  \bibinfo{author}{\bibfnamefont{S.~J.} \bibnamefont{Pollock}},
  \bibnamefont{and} \bibinfo{author}{\bibfnamefont{K.~K.}
  \bibnamefont{Perkins}}, \emph{\bibinfo{title}{Observations on student
  difficulties with mathematics in upper-division electricity and magnetism}},
  \bibinfo{journal}{Phys. Rev. ST Phys. Educ. Res.}
  \textbf{\bibinfo{volume}{8}}, \bibinfo{pages}{010111}
  (\bibinfo{year}{2012}{\natexlab{a}}).

\bibitem[{\citenamefont{Christensen and
  Thompson}(2012)}]{PhysRevSTPER.8.023101}
\bibinfo{author}{\bibfnamefont{W.~M.} \bibnamefont{Christensen}}
  \bibnamefont{and} \bibinfo{author}{\bibfnamefont{J.~R.}
  \bibnamefont{Thompson}}, \emph{\bibinfo{title}{Investigating graphical
  representations of slope and derivative without a physics context}},
  \bibinfo{journal}{Phys. Rev. ST Phys. Educ. Res.}
  \textbf{\bibinfo{volume}{8}}, \bibinfo{pages}{023101} (\bibinfo{year}{2012}).

\bibitem[{\citenamefont{Ambrose}(2004)}]{Ambrose:2004ed}
\bibinfo{author}{\bibfnamefont{B.~S.} \bibnamefont{Ambrose}},
  \emph{\bibinfo{title}{{Investigating student understanding in intermediate
  mechanics: Identifying the need for a tutorial approach to instruction}}},
  \bibinfo{journal}{Am. J. Phys.} \textbf{\bibinfo{volume}{72}},
  \bibinfo{pages}{453} (\bibinfo{year}{2004}).

\bibitem[{\citenamefont{Smith et~al.}(2010)\citenamefont{Smith, Thompson, and
  Mountcastle}}]{Smith:2010wx}
\bibinfo{author}{\bibfnamefont{T.}~\bibnamefont{Smith}},
  \bibinfo{author}{\bibfnamefont{J.}~\bibnamefont{Thompson}}, \bibnamefont{and}
  \bibinfo{author}{\bibfnamefont{D.}~\bibnamefont{Mountcastle}},
  \emph{\bibinfo{title}{{Addressing student difficulties with statistical
  mechanics: The Boltzmann factor}}}, \bibinfo{journal}{PERC 2010 Proceedings}
  p. \bibinfo{pages}{305} (\bibinfo{year}{2010}).

\bibitem[{\citenamefont{Pollock et~al.}(2007)\citenamefont{Pollock, Thompson,
  and Mountcastle}}]{Pollock:2007wv}
\bibinfo{author}{\bibfnamefont{E.~B.} \bibnamefont{Pollock}},
  \bibinfo{author}{\bibfnamefont{J.}~\bibnamefont{Thompson}}, \bibnamefont{and}
  \bibinfo{author}{\bibfnamefont{D.}~\bibnamefont{Mountcastle}},
  \emph{\bibinfo{title}{{Student understanding of the physics and mathematics
  of process variables in P-V diagrams}}}, \bibinfo{journal}{PERC 2007
  Proceedings} p. \bibinfo{pages}{168} (\bibinfo{year}{2007}).

\bibitem[{\citenamefont{Mason and Singh}(2010)}]{Mason:2010jd}
\bibinfo{author}{\bibfnamefont{A.}~\bibnamefont{Mason}} \bibnamefont{and}
  \bibinfo{author}{\bibfnamefont{C.}~\bibnamefont{Singh}},
  \emph{\bibinfo{title}{{Do advanced physics students learn from their mistakes
  without explicit intervention?}}}, \bibinfo{journal}{Am. J. Phys.}
  \textbf{\bibinfo{volume}{78}}, \bibinfo{pages}{760} (\bibinfo{year}{2010}).

\bibitem[{\citenamefont{Smith et~al.}(2009)\citenamefont{Smith, Christensen,
  and Thompson}}]{Smith2009entropy}
\bibinfo{author}{\bibfnamefont{T.~I.} \bibnamefont{Smith}},
  \bibinfo{author}{\bibfnamefont{W.~M.} \bibnamefont{Christensen}},
  \bibnamefont{and} \bibinfo{author}{\bibfnamefont{J.~R.}
  \bibnamefont{Thompson}}, \emph{\bibinfo{title}{Addressing student
  difficulties with concepts related to entropy, heat engines and the carnot
  cycle}}, \bibinfo{journal}{PERC 2009 Proceedings} p. \bibinfo{pages}{277}
  (\bibinfo{year}{2009}).

\bibitem[{\citenamefont{Meltzer}(2009)}]{Meltzer2009thermal}
\bibinfo{author}{\bibfnamefont{D.~E.} \bibnamefont{Meltzer}},
  \emph{\bibinfo{title}{Observations of general learning patterns in an
  upper‐level thermal physics course}}, \bibinfo{journal}{PERC 2009
  Proceedings} p.~\bibinfo{pages}{31} (\bibinfo{year}{2009}).

\bibitem[{\citenamefont{Michelini and Zuccarini}(2015)}]{Michelini2015}
\bibinfo{author}{\bibfnamefont{M.}~\bibnamefont{Michelini}} \bibnamefont{and}
  \bibinfo{author}{\bibfnamefont{G.}~\bibnamefont{Zuccarini}},
  \emph{\bibinfo{title}{{University Students' Reasoning on Physical Information
  Encoded in Quantum State at a Point in Time}}}, \bibinfo{journal}{PERC 2014
  Proceedings} p. \bibinfo{pages}{187} (\bibinfo{year}{2015}).

\bibitem[{\citenamefont{Sayre et~al.}(2007)\citenamefont{Sayre, Wittmann, and
  Donovan}}]{10.1063/1.2508697}
\bibinfo{author}{\bibfnamefont{E.~C.} \bibnamefont{Sayre}},
  \bibinfo{author}{\bibfnamefont{M.~C.} \bibnamefont{Wittmann}},
  \bibnamefont{and} \bibinfo{author}{\bibfnamefont{J.~E.}
  \bibnamefont{Donovan}}, \emph{\bibinfo{title}{Resource plasticity: Detailing
  a common chain of reasoning with damped harmonic motion}},
  \bibinfo{journal}{PERC 2006 Proceedings} p.~\bibinfo{pages}{85}
  (\bibinfo{year}{2007}).

\bibitem[{\citenamefont{Black and Wittmann}(2007)}]{10.1063/1.2820945}
\bibinfo{author}{\bibfnamefont{K.~E.} \bibnamefont{Black}} \bibnamefont{and}
  \bibinfo{author}{\bibfnamefont{M.~C.} \bibnamefont{Wittmann}},
  \emph{\bibinfo{title}{Epistemic games in integration: Modeling resource
  choice}}, \bibinfo{journal}{PERC 2007 Proceedings} p.~\bibinfo{pages}{53}
  (\bibinfo{year}{2007}).

\bibitem[{\citenamefont{Black and Wittmann}(2009)}]{Black:2009tu}
\bibinfo{author}{\bibfnamefont{K.~E.} \bibnamefont{Black}} \bibnamefont{and}
  \bibinfo{author}{\bibfnamefont{M.~C.} \bibnamefont{Wittmann}},
  \emph{\bibinfo{title}{{Procedural resource creation in intermediate
  mechanics}}}, \bibinfo{journal}{PERC 2009 Proceedings} p.~\bibinfo{pages}{97}
  (\bibinfo{year}{2009}).

\bibitem[{\citenamefont{Irving et~al.}(2013)\citenamefont{Irving, Martinuk, and
  Sayre}}]{Irving:2013bi}
\bibinfo{author}{\bibfnamefont{P.}~\bibnamefont{Irving}},
  \bibinfo{author}{\bibfnamefont{M.}~\bibnamefont{Martinuk}}, \bibnamefont{and}
  \bibinfo{author}{\bibfnamefont{E.}~\bibnamefont{Sayre}},
  \emph{\bibinfo{title}{{Transitions in students' epistemic framing along two
  axes}}}, \bibinfo{journal}{Phys. Rev. ST Phys. Educ. Res.}
  \textbf{\bibinfo{volume}{9}}, \bibinfo{pages}{010111} (\bibinfo{year}{2013}).

\bibitem[{\citenamefont{Wilcox et~al.}(2013{\natexlab{a}})\citenamefont{Wilcox,
  Caballero, Rehn, and Pollock}}]{PhysRevSTPER.9.020119}
\bibinfo{author}{\bibfnamefont{B.~R.} \bibnamefont{Wilcox}},
  \bibinfo{author}{\bibfnamefont{M.~D.} \bibnamefont{Caballero}},
  \bibinfo{author}{\bibfnamefont{D.~A.} \bibnamefont{Rehn}}, \bibnamefont{and}
  \bibinfo{author}{\bibfnamefont{S.~J.} \bibnamefont{Pollock}},
  \emph{\bibinfo{title}{Analytic framework for students' use of mathematics in
  upper-division physics}}, \bibinfo{journal}{Phys. Rev. ST Phys. Educ. Res.}
  \textbf{\bibinfo{volume}{9}}, \bibinfo{pages}{020119}
  (\bibinfo{year}{2013}{\natexlab{a}}).

\bibitem[{\citenamefont{Uhden et~al.}(2012)\citenamefont{Uhden, Karam,
  Pietrocola, and Pospiech}}]{Uhden2012}
\bibinfo{author}{\bibfnamefont{O.}~\bibnamefont{Uhden}},
  \bibinfo{author}{\bibfnamefont{R.}~\bibnamefont{Karam}},
  \bibinfo{author}{\bibfnamefont{M.}~\bibnamefont{Pietrocola}},
  \bibnamefont{and} \bibinfo{author}{\bibfnamefont{G.}~\bibnamefont{Pospiech}},
  \emph{\bibinfo{title}{{Modelling Mathematical Reasoning in Physics
  Education}}}, \bibinfo{journal}{Sci \& Educ} \textbf{\bibinfo{volume}{21}},
  \bibinfo{pages}{485} (\bibinfo{year}{2012}).

\bibitem[{\citenamefont{{L{\'o}pez-Gay, R and S{\'a}ez, J Mart{\'\i}nez and
  Torregrosa, J Mart{\'\i}nez}}(2015)}]{Lopez2015}
\bibinfo{author}{\bibnamefont{{L{\'o}pez-Gay, R and S{\'a}ez, J Mart{\'\i}nez
  and Torregrosa, J Mart{\'\i}nez}}}, \emph{\bibinfo{title}{{Obstacles to
  Mathematization in Physics: The Case of the Differential}}},
  \bibinfo{journal}{Sci \& Educ} \textbf{\bibinfo{volume}{24}},
  \bibinfo{pages}{1} (\bibinfo{year}{2015}).

\bibitem[{\citenamefont{Michelsen}(2006)}]{Michelsen2006}
\bibinfo{author}{\bibfnamefont{C.}~\bibnamefont{Michelsen}},
  \emph{\bibinfo{title}{Functions: a modelling tool in mathematics and
  science}}, \bibinfo{journal}{ZDM} \textbf{\bibinfo{volume}{38}},
  \bibinfo{pages}{269} (\bibinfo{year}{2006}).

\bibitem[{\citenamefont{Andreas~Redfors and Juter}(2013)}]{Redfors2013}
\bibinfo{author}{\bibfnamefont{{\..\O}.~H.} \bibnamefont{Andreas~Redfors},
  \bibfnamefont{Lena~Hansson}} \bibnamefont{and}
  \bibinfo{author}{\bibfnamefont{K.}~\bibnamefont{Juter}},
  \emph{\bibinfo{title}{{The Role of Mathematics in the Teaching and Learning
  of Physics}}}, \bibinfo{journal}{ESERA 2013 Conference Proceedings} p.
  \bibinfo{pages}{376} (\bibinfo{year}{2013}).

\bibitem[{\citenamefont{Karam}(2014)}]{Karam2014}
\bibinfo{author}{\bibfnamefont{R.}~\bibnamefont{Karam}},
  \emph{\bibinfo{title}{Framing the structural role of mathematics in physics
  lectures: A case study on electromagnetism}}, \bibinfo{journal}{Phys. Rev. ST
  Phys. Educ. Res.} \textbf{\bibinfo{volume}{10}}, \bibinfo{pages}{010119}
  (\bibinfo{year}{2014}).

\bibitem[{\citenamefont{Redish}(2006)}]{Redish:2006un}
\bibinfo{author}{\bibfnamefont{E.}~\bibnamefont{Redish}},
  \emph{\bibinfo{title}{{Problem solving and the use of math in physics
  courses}}}, \bibinfo{journal}{Arxiv preprint physics/0608268}
  (\bibinfo{year}{2006}).

\bibitem[{\citenamefont{Hammer}(2000)}]{Hammer:2000vf}
\bibinfo{author}{\bibfnamefont{D.}~\bibnamefont{Hammer}},
  \emph{\bibinfo{title}{{Student resources for learning introductory
  physics}}}, \bibinfo{journal}{Am. J. Phys.} \textbf{\bibinfo{volume}{68}},
  \bibinfo{pages}{S52} (\bibinfo{year}{2000}).

\bibitem[{\citenamefont{Collins and Ferguson}(1993)}]{Collins:1993vz}
\bibinfo{author}{\bibfnamefont{A.}~\bibnamefont{Collins}} \bibnamefont{and}
  \bibinfo{author}{\bibfnamefont{W.}~\bibnamefont{Ferguson}},
  \emph{\bibinfo{title}{{Epistemic forms and epistemic games: Structures and
  strategies to guide inquiry}}}, \bibinfo{journal}{Educational Psychologist}
  \textbf{\bibinfo{volume}{28}}, \bibinfo{pages}{25} (\bibinfo{year}{1993}).

\bibitem[{\citenamefont{Hammer et~al.}(2005)\citenamefont{Hammer, Elby, Scherr,
  and Redish}}]{Hammer:2005th}
\bibinfo{author}{\bibfnamefont{D.}~\bibnamefont{Hammer}},
  \bibinfo{author}{\bibfnamefont{A.}~\bibnamefont{Elby}},
  \bibinfo{author}{\bibfnamefont{R.}~\bibnamefont{Scherr}}, \bibnamefont{and}
  \bibinfo{author}{\bibfnamefont{E.}~\bibnamefont{Redish}}, in
  \emph{\bibinfo{booktitle}{Transfer of learning from a modern
  multidisciplinary perspective}}, edited by
  \bibinfo{editor}{\bibfnamefont{J.~P.} \bibnamefont{Mestre}}
  (\bibinfo{publisher}{IAP}, \bibinfo{year}{2005}).

\bibitem[{\citenamefont{Arons}(1997)}]{arons1997teaching}
\bibinfo{author}{\bibfnamefont{A.}~\bibnamefont{Arons}},
  \emph{\bibinfo{title}{Teaching introductory physics}}
  (\bibinfo{publisher}{Wiley}, \bibinfo{year}{1997}).

\bibitem[{\citenamefont{Close and Heron}(2011)}]{heron2011angmom}
\bibinfo{author}{\bibfnamefont{H.}~\bibnamefont{Close}} \bibnamefont{and}
  \bibinfo{author}{\bibfnamefont{P.}~\bibnamefont{Heron}},
  \emph{\bibinfo{title}{Student understanding of the angular momentum of
  classical particles}}, \bibinfo{journal}{Am. J. Phys.}
  \textbf{\bibinfo{volume}{79}}, \bibinfo{pages}{1068} (\bibinfo{year}{2011}).

\bibitem[{\citenamefont{Lindsey et~al.}(2012)\citenamefont{Lindsey, Heron, and
  Shaffer}}]{Lindsey:2012vj}
\bibinfo{author}{\bibfnamefont{B.}~\bibnamefont{Lindsey}},
  \bibinfo{author}{\bibfnamefont{P.}~\bibnamefont{Heron}}, \bibnamefont{and}
  \bibinfo{author}{\bibfnamefont{P.}~\bibnamefont{Shaffer}},
  \emph{\bibinfo{title}{Student understanding of energy: Difficulties related
  to systems}}, \bibinfo{journal}{Am. J. Phys.} \textbf{\bibinfo{volume}{80}},
  \bibinfo{pages}{154} (\bibinfo{year}{2012}).

\bibitem[{\citenamefont{Ortiz et~al.}(2005)\citenamefont{Ortiz, Heron, and
  Shaffer}}]{Ortiz:2005hu}
\bibinfo{author}{\bibfnamefont{L.~G.} \bibnamefont{Ortiz}},
  \bibinfo{author}{\bibfnamefont{P.~R.~L.} \bibnamefont{Heron}},
  \bibnamefont{and} \bibinfo{author}{\bibfnamefont{P.~S.}
  \bibnamefont{Shaffer}}, \emph{\bibinfo{title}{{Student understanding of
  static equilibrium: Predicting and accounting for balancing}}},
  \bibinfo{journal}{Am. J. Phys.} \textbf{\bibinfo{volume}{73}},
  \bibinfo{pages}{545} (\bibinfo{year}{2005}).

\bibitem[{\citenamefont{Pollock}(2007)}]{pollock2007longitudinal}
\bibinfo{author}{\bibfnamefont{S.}~\bibnamefont{Pollock}},
  \emph{\bibinfo{title}{{A Longitudinal Study of the Impact of Curriculum on
  Conceptual Understanding in E\&M}}}, \bibinfo{journal}{PERC 2007 Proceedings}
  pp. \bibinfo{pages}{172--175} (\bibinfo{year}{2007}).

\bibitem[{\citenamefont{Finkelstein and Pollock}(2005)}]{finkelstein2005tip}
\bibinfo{author}{\bibfnamefont{N.}~\bibnamefont{Finkelstein}} \bibnamefont{and}
  \bibinfo{author}{\bibfnamefont{S.}~\bibnamefont{Pollock}},
  \emph{\bibinfo{title}{Replicating and understanding successful innovations:
  Implementing tutorials in introductory physics}}, \bibinfo{journal}{Phys.
  Rev. ST Phys. Educ. Res.} \textbf{\bibinfo{volume}{1}},
  \bibinfo{pages}{010101} (\bibinfo{year}{2005}).

\bibitem[{\citenamefont{Sadaghiani}(2005)}]{sadaghiani2005qm}
\bibinfo{author}{\bibfnamefont{H.}~\bibnamefont{Sadaghiani}},
  \bibinfo{type}{Ph.d.}, \bibinfo{school}{The Ohio State University}
  (\bibinfo{year}{2005}).

\bibitem[{\citenamefont{Loverude}(2009)}]{loverude2009prob}
\bibinfo{author}{\bibfnamefont{M.}~\bibnamefont{Loverude}},
  \emph{\bibinfo{title}{Student understanding of basic probability concepts in
  an upper-division thermal physics course}}, \bibinfo{journal}{PERC 2009
  Proceedings} pp. \bibinfo{pages}{189--192} (\bibinfo{year}{2009}).

\bibitem[{\citenamefont{Loverude and Li}(2014)}]{loverude2013mm}
\bibinfo{author}{\bibfnamefont{M.}~\bibnamefont{Loverude}} \bibnamefont{and}
  \bibinfo{author}{\bibfnamefont{S.}~\bibnamefont{Li}},
  \emph{\bibinfo{title}{{"Surprisingly, there is an actual physical
  application…" Student understanding in Math Methods}}},
  \bibinfo{journal}{PERC 2013 Proceedings} p. \bibinfo{pages}{229}
  (\bibinfo{year}{2014}).

\bibitem[{\citenamefont{Bucy et~al.}(2007)\citenamefont{Bucy, Thompson, and
  Mountcastle}}]{bucy2006pDerivs}
\bibinfo{author}{\bibfnamefont{B.}~\bibnamefont{Bucy}},
  \bibinfo{author}{\bibfnamefont{J.}~\bibnamefont{Thompson}}, \bibnamefont{and}
  \bibinfo{author}{\bibfnamefont{D.}~\bibnamefont{Mountcastle}},
  \emph{\bibinfo{title}{{Student (Mis)application of Partial Differentiation to
  Material Properties}}}, \bibinfo{journal}{PERC 2006 Proceedings} pp.
  \bibinfo{pages}{157--160} (\bibinfo{year}{2007}).

\bibitem[{\citenamefont{Thompson et~al.}(2006)\citenamefont{Thompson, Bucy, and
  Mountcastle}}]{thompson2005pDerivs}
\bibinfo{author}{\bibfnamefont{J.}~\bibnamefont{Thompson}},
  \bibinfo{author}{\bibfnamefont{B.}~\bibnamefont{Bucy}}, \bibnamefont{and}
  \bibinfo{author}{\bibfnamefont{D.}~\bibnamefont{Mountcastle}},
  \emph{\bibinfo{title}{Assessing student understanding of partial derivatives
  in thermodynamics}}, \bibinfo{journal}{PERC 2005 Proceedings}
  \textbf{\bibinfo{volume}{818}}, \bibinfo{pages}{77} (\bibinfo{year}{2006}).

\bibitem[{\citenamefont{Louca et~al.}(2004)\citenamefont{Louca, Elby, Hammer,
  and Kagey}}]{Louca2004}
\bibinfo{author}{\bibfnamefont{L.}~\bibnamefont{Louca}},
  \bibinfo{author}{\bibfnamefont{A.}~\bibnamefont{Elby}},
  \bibinfo{author}{\bibfnamefont{D.}~\bibnamefont{Hammer}}, \bibnamefont{and}
  \bibinfo{author}{\bibfnamefont{T.}~\bibnamefont{Kagey}},
  \emph{\bibinfo{title}{Epistemological resources: Applying a new
  epistemological framework to science instruction}},
  \bibinfo{journal}{Educational Psychologist} \textbf{\bibinfo{volume}{39}},
  \bibinfo{pages}{57} (\bibinfo{year}{2004}).

\bibitem[{\citenamefont{Elby and Hammer}(2001)}]{Elby2001b}
\bibinfo{author}{\bibfnamefont{A.}~\bibnamefont{Elby}} \bibnamefont{and}
  \bibinfo{author}{\bibfnamefont{D.}~\bibnamefont{Hammer}},
  \emph{\bibinfo{title}{On the substance of a sophisticated epistemology}},
  \bibinfo{journal}{Science Education} \textbf{\bibinfo{volume}{85}},
  \bibinfo{pages}{554} (\bibinfo{year}{2001}).

\bibitem[{\citenamefont{Tuminaro and Redish}(2007)}]{Tuminaro:2007hr}
\bibinfo{author}{\bibfnamefont{J.}~\bibnamefont{Tuminaro}} \bibnamefont{and}
  \bibinfo{author}{\bibfnamefont{E.}~\bibnamefont{Redish}},
  \emph{\bibinfo{title}{{Elements of a cognitive model of physics problem
  solving: Epistemic games}}}, \bibinfo{journal}{Phys. Rev. ST Phys. Ed. Res.}
  \textbf{\bibinfo{volume}{3}}, \bibinfo{pages}{020101} (\bibinfo{year}{2007}).

\bibitem[{\citenamefont{Scherr and Hammer}(2009)}]{Scherr:2009gm}
\bibinfo{author}{\bibfnamefont{R.~E.} \bibnamefont{Scherr}} \bibnamefont{and}
  \bibinfo{author}{\bibfnamefont{D.}~\bibnamefont{Hammer}},
  \emph{\bibinfo{title}{{Student Behavior and Epistemological Framing: Examples
  from Collaborative Active-Learning Activities in Physics}}},
  \bibinfo{journal}{Cognition and Instruction} \textbf{\bibinfo{volume}{27}},
  \bibinfo{pages}{147} (\bibinfo{year}{2009}).

\bibitem[{\citenamefont{Elby}(2001)}]{Elby2001a}
\bibinfo{author}{\bibfnamefont{A.}~\bibnamefont{Elby}},
  \emph{\bibinfo{title}{Helping physics students learn how to learn}},
  \bibinfo{journal}{Am. J. Phys.} \textbf{\bibinfo{volume}{69}},
  \bibinfo{pages}{S54} (\bibinfo{year}{2001}).

\bibitem[{\citenamefont{Sabella and Redish}(2007)}]{Sabella:2007ff}
\bibinfo{author}{\bibfnamefont{M.~S.} \bibnamefont{Sabella}} \bibnamefont{and}
  \bibinfo{author}{\bibfnamefont{E.~F.} \bibnamefont{Redish}},
  \emph{\bibinfo{title}{{Knowledge organization and activation in physics
  problem solving}}}, \bibinfo{journal}{Am. J. Phys.}
  \textbf{\bibinfo{volume}{75}}, \bibinfo{pages}{1017} (\bibinfo{year}{2007}).

\bibitem[{\citenamefont{Bing and Redish}(2009)}]{Bing:2009gq}
\bibinfo{author}{\bibfnamefont{T.}~\bibnamefont{Bing}} \bibnamefont{and}
  \bibinfo{author}{\bibfnamefont{E.}~\bibnamefont{Redish}},
  \emph{\bibinfo{title}{{Analyzing problem solving using math in physics:
  Epistemological framing via warrants}}}, \bibinfo{journal}{Phys. Rev. ST
  Phys. Educ. Res.} \textbf{\bibinfo{volume}{5}}, \bibinfo{pages}{020108}
  (\bibinfo{year}{2009}).

\bibitem[{\citenamefont{Kustusch et~al.}(2014)\citenamefont{Kustusch, Roundy,
  Dray, and Manogue}}]{Kustusch:2014gz}
\bibinfo{author}{\bibfnamefont{M.}~\bibnamefont{Kustusch}},
  \bibinfo{author}{\bibfnamefont{D.}~\bibnamefont{Roundy}},
  \bibinfo{author}{\bibfnamefont{T.}~\bibnamefont{Dray}}, \bibnamefont{and}
  \bibinfo{author}{\bibfnamefont{C.}~\bibnamefont{Manogue}},
  \emph{\bibinfo{title}{{Partial derivative games in thermodynamics: A
  cognitive task analysis}}}, \bibinfo{journal}{Phys. Rev. ST Phys. Educ. Res.}
  \textbf{\bibinfo{volume}{10}}, \bibinfo{pages}{010101}
  (\bibinfo{year}{2014}).

\bibitem[{\citenamefont{Redish}(2004)}]{Redish2004}
\bibinfo{author}{\bibfnamefont{E.~F.} \bibnamefont{Redish}},
  \emph{\bibinfo{title}{A theoretical framework for physics education research:
  Modeling student thinking}}, \bibinfo{journal}{arXiv preprint
  physics/0411149}  (\bibinfo{year}{2004}).

\bibitem[{\citenamefont{diSessa}(1988)}]{diSessa1988}
\bibinfo{author}{\bibfnamefont{A.~A.} \bibnamefont{diSessa}},
  \emph{\bibinfo{title}{Constructivism in the Computer Age}}
  (\bibinfo{publisher}{Lawrence Erlbaum}, \bibinfo{address}{Hillsdale, NJ},
  \bibinfo{year}{1988}), chap. \bibinfo{chapter}{Knowledge in pieces}, pp.
  \bibinfo{pages}{1--24}.

\bibitem[{\citenamefont{Sherin}(2001)}]{Sherin:2001wq}
\bibinfo{author}{\bibfnamefont{B.}~\bibnamefont{Sherin}},
  \emph{\bibinfo{title}{{How students understand physics equations}}},
  \bibinfo{journal}{Cognition and Instruction} \textbf{\bibinfo{volume}{19}},
  \bibinfo{pages}{479} (\bibinfo{year}{2001}).

\bibitem[{\citenamefont{Chasteen et~al.}(2009)\citenamefont{Chasteen, Perkins,
  Beale, Pollock, and Wieman}}]{Chasteen:2009wn}
\bibinfo{author}{\bibfnamefont{S.}~\bibnamefont{Chasteen}},
  \bibinfo{author}{\bibfnamefont{K.}~\bibnamefont{Perkins}},
  \bibinfo{author}{\bibfnamefont{P.}~\bibnamefont{Beale}},
  \bibinfo{author}{\bibfnamefont{S.}~\bibnamefont{Pollock}}, \bibnamefont{and}
  \bibinfo{author}{\bibfnamefont{C.}~\bibnamefont{Wieman}},
  \emph{\bibinfo{title}{{A Thoughtful Approach to Instruction: Course
  transformation for the rest of us}}}, \bibinfo{journal}{J. Coll. Sci. Teach.}
  \textbf{\bibinfo{volume}{40}}, \bibinfo{pages}{70} (\bibinfo{year}{2009}).

\bibitem[{\citenamefont{Baily et~al.}(2013)\citenamefont{Baily, Dubson, and
  Pollock}}]{2012arXiv1207.6040B}
\bibinfo{author}{\bibfnamefont{C.}~\bibnamefont{Baily}},
  \bibinfo{author}{\bibfnamefont{M.}~\bibnamefont{Dubson}}, \bibnamefont{and}
  \bibinfo{author}{\bibfnamefont{S.~J.} \bibnamefont{Pollock}},
  \emph{\bibinfo{title}{{Research-Based Course Materials and Assessments for
  Upper-Division Electrodynamics (E{\&}M II)}}}, \bibinfo{journal}{PERC 2012
  Proceedings}  (\bibinfo{year}{2013}).

\bibitem[{\citenamefont{Goldhaber et~al.}(2009)\citenamefont{Goldhaber,
  Pollock, Dubson, Beale, and Perkins}}]{goldhaber2009qmat}
\bibinfo{author}{\bibfnamefont{S.}~\bibnamefont{Goldhaber}},
  \bibinfo{author}{\bibfnamefont{S.}~\bibnamefont{Pollock}},
  \bibinfo{author}{\bibfnamefont{M.}~\bibnamefont{Dubson}},
  \bibinfo{author}{\bibfnamefont{P.}~\bibnamefont{Beale}}, \bibnamefont{and}
  \bibinfo{author}{\bibfnamefont{K.}~\bibnamefont{Perkins}},
  \emph{\bibinfo{title}{{Transforming Upper-Division Quantum Mechanics:
  Learning Goals and Assessment}}}, \bibinfo{journal}{PERC 2009 Proceedings} p.
  \bibinfo{pages}{145} (\bibinfo{year}{2009}).

\bibitem[{\citenamefont{Pollock et~al.}(2012)\citenamefont{Pollock, Pepper, and
  Marino}}]{Pollock:2012uy}
\bibinfo{author}{\bibfnamefont{S.}~\bibnamefont{Pollock}},
  \bibinfo{author}{\bibfnamefont{R.}~\bibnamefont{Pepper}}, \bibnamefont{and}
  \bibinfo{author}{\bibfnamefont{A.~D.} \bibnamefont{Marino}},
  \emph{\bibinfo{title}{{Issues and progress in transforming a middle-division
  classical mechanics/math methods course}}}, \bibinfo{journal}{PERC 2011
  Proceedings} p. \bibinfo{pages}{303} (\bibinfo{year}{2012}).

\bibitem[{\citenamefont{Chasteen et~al.}(2012)\citenamefont{Chasteen, Pepper,
  Pollock, Perkins, Rebello, Engelhardt, and Singh}}]{Chasteen:2012vb}
\bibinfo{author}{\bibfnamefont{S.}~\bibnamefont{Chasteen}},
  \bibinfo{author}{\bibfnamefont{R.}~\bibnamefont{Pepper}},
  \bibinfo{author}{\bibfnamefont{S.}~\bibnamefont{Pollock}},
  \bibinfo{author}{\bibfnamefont{K.}~\bibnamefont{Perkins}},
  \bibinfo{author}{\bibfnamefont{N.}~\bibnamefont{Rebello}},
  \bibinfo{author}{\bibfnamefont{P.}~\bibnamefont{Engelhardt}},
  \bibnamefont{and} \bibinfo{author}{\bibfnamefont{C.}~\bibnamefont{Singh}},
  \emph{\bibinfo{title}{{But does it last? Sustaining a research-based
  curriculum in upper-division electricity {\&} magnetism}}},
  \bibinfo{journal}{PERC 2011 Proceedings} p. \bibinfo{pages}{139}
  (\bibinfo{year}{2012}).

\bibitem[{\citenamefont{Pepper et~al.}(2012{\natexlab{b}})\citenamefont{Pepper,
  Chasteen, Pollock, and Perkins}}]{2012AIPC.1413..291P}
\bibinfo{author}{\bibfnamefont{R.~E.} \bibnamefont{Pepper}},
  \bibinfo{author}{\bibfnamefont{S.~V.} \bibnamefont{Chasteen}},
  \bibinfo{author}{\bibfnamefont{S.~J.} \bibnamefont{Pollock}},
  \bibnamefont{and} \bibinfo{author}{\bibfnamefont{K.~K.}
  \bibnamefont{Perkins}}, \emph{\bibinfo{title}{{Facilitating faculty
  conversations: Development of consensus learning goals}}},
  \bibinfo{journal}{PERC 2011 Proceedings} p.~\bibinfo{pages}{29}
  (\bibinfo{year}{2012}{\natexlab{b}}).

\bibitem[{\citenamefont{Wilcox and Pollock}(2015)}]{2014arXiv1407.6311W}
\bibinfo{author}{\bibfnamefont{B.~R.} \bibnamefont{Wilcox}} \bibnamefont{and}
  \bibinfo{author}{\bibfnamefont{S.~J.} \bibnamefont{Pollock}},
  \emph{\bibinfo{title}{Upper-division student difficulties with the dirac
  delta function}}, \bibinfo{journal}{Phys. Rev. ST Phys. Educ. Res.}
  \textbf{\bibinfo{volume}{11}}, \bibinfo{pages}{010108}
  (\bibinfo{year}{2015}).

\bibitem[{\citenamefont{Wright and Williams}(1986)}]{wright1986wise}
\bibinfo{author}{\bibfnamefont{D.~S.} \bibnamefont{Wright}} \bibnamefont{and}
  \bibinfo{author}{\bibfnamefont{C.~D.} \bibnamefont{Williams}},
  \emph{\bibinfo{title}{A wise strategy for introductory physics}},
  \bibinfo{journal}{The Physics Teacher} \textbf{\bibinfo{volume}{24}},
  \bibinfo{pages}{020105} (\bibinfo{year}{1986}).

\bibitem[{\citenamefont{Heller et~al.}(1992)\citenamefont{Heller, Keith, and
  Anderson}}]{1992AmJPh..60..627H}
\bibinfo{author}{\bibfnamefont{P.}~\bibnamefont{Heller}},
  \bibinfo{author}{\bibfnamefont{R.}~\bibnamefont{Keith}}, \bibnamefont{and}
  \bibinfo{author}{\bibfnamefont{S.}~\bibnamefont{Anderson}},
  \emph{\bibinfo{title}{{Teaching problem solving through cooperative grouping.
  Part 1: Group versus individual problem solving}}}, \bibinfo{journal}{Am. J.
  Phys.} \textbf{\bibinfo{volume}{60}}, \bibinfo{pages}{627}
  (\bibinfo{year}{1992}).

\bibitem[{\citenamefont{Catrambone}(2011)}]{Catrambone:vf}
\bibinfo{author}{\bibfnamefont{R.}~\bibnamefont{Catrambone}},
  \emph{\bibinfo{title}{{Task Analysis by Problem Solving (TAPS): Uncovering
  Expert Knowledge to Develop High-Quality Instructional Materials and
  Training}}}, \bibinfo{journal}{2011 Learning and Technology Symposium}
  (\bibinfo{year}{2011}).

\bibitem[{\citenamefont{Caballero et~al.}(2013)\citenamefont{Caballero, Wilcox,
  Pepper, and Pollock}}]{Caballero:2012wr}
\bibinfo{author}{\bibfnamefont{M.~D.} \bibnamefont{Caballero}},
  \bibinfo{author}{\bibfnamefont{B.~R.} \bibnamefont{Wilcox}},
  \bibinfo{author}{\bibfnamefont{R.~E.} \bibnamefont{Pepper}},
  \bibnamefont{and} \bibinfo{author}{\bibfnamefont{S.~J.}
  \bibnamefont{Pollock}}, \emph{\bibinfo{title}{{ACER: A Framework on the Use
  of Mathematics in Upper-division Physics}}}, \bibinfo{journal}{PERC 2012
  Proceedings} p.~\bibinfo{pages}{90} (\bibinfo{year}{2013}).

\bibitem[{\citenamefont{Wilcox et~al.}(2013{\natexlab{b}})\citenamefont{Wilcox,
  Caballero, Pepper, and Pollock}}]{2012arXiv1207.1283W}
\bibinfo{author}{\bibfnamefont{B.~R.} \bibnamefont{Wilcox}},
  \bibinfo{author}{\bibfnamefont{M.~D.} \bibnamefont{Caballero}},
  \bibinfo{author}{\bibfnamefont{R.~E.} \bibnamefont{Pepper}},
  \bibnamefont{and} \bibinfo{author}{\bibfnamefont{S.~J.}
  \bibnamefont{Pollock}}, \emph{\bibinfo{title}{{Upper-division Student
  Understanding of Coulomb's Law: Difficulties with Continuous Charge
  Distributions}}}, \bibinfo{journal}{PERC 2012 Proceedings} p.
  \bibinfo{pages}{418} (\bibinfo{year}{2013}{\natexlab{b}}).

\bibitem[{\citenamefont{Scherr}(2009)}]{scherr:2009dn}
\bibinfo{author}{\bibfnamefont{R.}~\bibnamefont{Scherr}},
  \emph{\bibinfo{title}{{Video analysis for insight and coding: Examples from
  tutorials in introductory physics}}}, \bibinfo{journal}{Phys. Rev. ST Phys.
  Educ. Res.} \textbf{\bibinfo{volume}{5}}, \bibinfo{pages}{020106}
  (\bibinfo{year}{2009}).

\bibitem[{CUm()}]{CUmaterials}
\emph{\bibinfo{title}{{Upper-Division Course Transformations}}},
  \bibinfo{howpublished}{\url{http://per.colorado.edu/sei}},
  \bibinfo{note}{accessed: 2014-09-23}.

\end{thebibliography}
\bibliographystyle{apsper}

\end{document}